\documentclass[11pt,a4paper]{article}
\usepackage{jheppub}
\usepackage{amsmath}
\usepackage{dsfont}
\usepackage{ulem}
\usepackage{natbib}
\usepackage{xcolor}
\usepackage[hang,flushmargin]{footmisc}
\usepackage{tikz-cd}

\makeatletter\renewcommand{\@biblabel}[1]{#1.}\makeatother

\newcommand{\mb}[1]{\mathbf{#1}}
\newcommand{\nn}{\nonumber}

\def\i{{\rm i}}
\def\e{{\rm e}}
\def\d{{\rm d}}

\def\I{{\textrm{I}}}
\def\II{{\textrm{II}}}

\preprint{
{\small{\textsf{}}}}

\title{$q$-Virasoro modular triple}

\author{Fabrizio Nieri, Yiwen Pan and Maxim Zabzine}

\affiliation{Department of Physics and Astronomy, Uppsala University,\\
Box 516, SE-75120 Uppsala, Sweden.}

\emailAdd{fb.nieri@gmail.com}
\emailAdd{yiwen.pan@physics.uu.se}
\emailAdd{maxim.zabzine@physics.uu.se}

\abstract{Inspired by 5d supersymmetric Yang-Mills theories placed on the compact space $\mathbb{S}^5$, we propose an intriguing algebraic construction for the $q$-Virasoro algebra. We show that, when multiple $q$-Virasoro ``chiral" sectors have to be fused together, a natural $\mathrm{SL}(3,\mathbb{Z})$ structure arises. This construction, which we call the modular triple, is consistent with the observed triple factorization properties of supersymmetric partition functions derived from localization arguments. We also give a 2d CFT-like construction of the modular triple, and conjecture for the first time a (non-local) Lagrangian formulation for a $q$-Virasoro model, resembling ordinary Liouville theory.
}

\keywords{Supersymmetric gauge theories, deformed Virasoro algebra, modular double.}

\begin{document}
%\today
\maketitle
%\newpage

\flushbottom

\section{Introduction}

The discovery of algebraic structures in quantum field theories has always led to dramatic improvements of our understanding thereof. The hidden Yangian symmetry of planar 4d $\mathcal{N}=4$ Yang-Mills \cite{Drummond:2009fd} and the Virasoro/W symmetry of 4d $\mathcal{N}=2$ class $\mathcal{S}$ theories \cite{Gaiotto:2009we}, a.k.a. the AGT correspondence \cite{Alday:2009aq}, are perhaps the most famous examples. In the former case, the infinite dimensional symmetry algebra points towards the complete integrability of the model and has helped in computing the exact S-matrix (see \textit{e.g.} \cite{Beisert:2010jr} for a review). In the latter case, the 2d CFT Liouville/Toda description of the gauge theory placed on the compact space $\mathbb{S}^4$ \cite{Pestun:2007rz} has allowed to establish a powerful dictionary between observables of the two sides. More generally, the circle of ideas around the BPS/CFT correspondence (see \textit{e.g.} \cite{Nekrasov:2015wsu} for a recent review) has led to an enormous amount of works relating supersymmetric gauge theories and (quantum) algebras (see \textit{e.g.} \cite{Nekrasov:2012xe,Nekrasov:2013xda}). For the purposes of this paper, the most interesting aspects are the identification of 5d Nekrasov partition functions on $\mathbb{R}^4_{q,t}\times \mathbb{S}^1$ \cite{Nekrasov:2002qd,Nekrasov:2003rj} with ``chiral" $q$-Virasoro/W correlators \cite{Awata:2009ur,Awata:2010yy,Awata:2011dc,Mironov:2011dk,Carlsson:2013jka,Aganagic:2013tta,Aganagic:2014oia,Aganagic:2014kja,Kimura:2015rgi,Mironov:2016yue,Bourgine:2016vsq,Bourgine:2017jsi}, as well as the extension of the AGT duality to 5d $\mathcal{N}=1$ supersymmetric gauge theories on compact spaces and $q$-deformed CFTs with multiple ``chiral" sectors \cite{Nieri:2013yra,Nieri:2013vba,Bao:2013pwa,Mitev:2014isa}. 

While the original 4d AGT setup is now quite well understood \cite{Cordova:2016cmu} thanks to the 6d $(2,0)$ SCFT construction of class $\mathcal{S}$ theories and the Lagrangian formulation of Liouville/Toda theory, the 5d picture is less developed. Among the main obstacles, we can recall that the origin of the $q$-deformation has been proposed to lie in the 6d $(2,0)$ Little String Theory \cite{Aganagic:2015cta}, much less under control than its conformal limit, and the lack of any Lagrangian $q$-CFT model with underlying $q$-Virasoro/W symmetry. In this paper, we will focus on the algebraic side of the BPS/CFT correspondence, trying to shed some new light on $q$-Virasoro systems by taking inspiration from the gauge theory results. 

Our strategy is to mimic as much as possible the structure of ordinary 2d CFTs, including the simplest real free boson theory. An important feature of these models is the holomorphic/anti-holomorphic factorization of physical (monodromy invariant) correlators, which is due to the existence of two chiral sectors (left and right moving) each associated to an independent copy of the Virasoro algebra. It is natural to expect a similar structure in the $q$-deformed case too, in which multiple $q$-Virasoro ``chiral" sectors are suitably glued together. Here it is where the gauge theory inspiration comes into play. Among all the possible pairings, the study of 5d supersymmetric gauge theories on $\mathbb{S}^5$ \cite{Lockhart:2012vp,Kallen:2012cs,Kallen:2012va,Qiu:2013aga,Kim:2012qf,Kim:2012ava,Nieri:2013vba,Qiu:2014oqa,Pan:2014bwa} (and/or co-dimension $2$ defects on $\mathbb{S}^3$ \cite{Pasquetti:2011fj,Beem:2012mb,Bullimore:2014awa})  suggests that one should consider $2$ or $3$ ``chiral" sectors glued together in an $\mathrm{SL}(2,\mathbb{Z})$ or $\mathrm{SL}(3,\mathbb{Z})$ fashion. In this case, it has also been shown that the resulting correlators are consistent with a $q$-deformed version of the bootstrap equations in Liouville theory \cite{Nieri:2013yra}. 

The $\mathrm{SL}(2,\mathbb{Z})$-pairing of \textit{two} $q$-Virasoro sectors has been studied in detail in \cite{Nedelin:2016gwu}, and the emerging algebraic structure has been given the name \textit{modular double}, in analogy with the $\mathcal{U}_q(\mathfrak{sl}_2)$ modular double introduced in \cite{Faddeev:1999fe,Ponsot:1999uf,Ponsot:2000mt}. In the case of major interest to us, the $\{q_i,t_i\}_{i=1,2}$ deformation parameters of the two commuting $q$-Virasoro algebras are related by the $S\in \mathrm{SL}(2,\mathbb{Z})$ element 
\begin{equation}
\begin{tikzcd}[row sep=tiny]
\big(q_1 = \e^{2 \pi \i \tau}, t_1=\e^{2\pi\i\sigma}\big) \arrow[mapsto]{r}{-S} & \big(q_2 = \e^{2\pi\i/\tau}, t_2=\e^{2\pi\i\sigma/\tau} \big) \ ,
\end{tikzcd}\nn
\end{equation}
and it has been shown that there exist two commuting sets of Ward Identities that correlators of the modular double must satisfy. The whole structure can be compactly encoded into the existence of a \textit{single screening current} $\mb{S}(X)$ defined by (roughly)
\begin{align}
\left[\mb{T}_{m,i},\mb{S}(X)\right]=\textrm{total $X$-difference}\;,\nn
\end{align}
where $\{\mb{T}_{m,i}\;, m\in\mathbb{Z}\}_{i=1,2}$ are two commuting copies of $q$-Virasoro generators. This is strongly reminiscent of what happens in ordinary 2d CFT, where a primary operator (screening current) $\mb{s}(z,\bar z)$ of conformal dimension $(1,1)$  satisfies
\begin{align}
\left[\mb{L}_{m},\mb{s}(z,\bar z)\right]=\textrm{total $z$-derivative},\quad \left[\overline{\mb{L}}_{m},\mb{s}(z,\bar z)\right]=\textrm{total $\bar z$-derivative}\;,\nn
\end{align}
where $\{\mb{L}_{m}\; m\in\mathbb{Z}\}$ and $\{\overline{\mb{L}}_{m}\; m\in\mathbb{Z}\}$ are the Virasoro generators of the two chiral sectors of the conformal algebra.  

One of the main results of this paper is that the above $\mathrm{SL}(2,\mathbb{Z})$-pairing can (in fact, must) be enhanced to  an $\mathrm{SL}(3,\mathbb{Z})$-pairing, in which $\textit{three}$ $q$-Virasoro ``chiral" sectors with deformation parameters $\{q_i,t_i\}_{i=1,2,3}$  are related as follows
\begin{equation}
\begin{tikzpicture}[commutative diagrams/every diagram]
  \node (P0) at (0,1) {$\big(q_1 = \e^{2 \pi \i \tau}, t_1=\e^{2\pi\i\sigma}\big)$};
  \node (P1) at (-4,-1) {$\big(q_2 = \e^{-2 \pi \i \sigma/\tau},t_2=\e^{-2\pi\i/\tau}\big)$} ;
  \node (P2) at (4,-1) {$\big(q_3 = \e^{-2 \pi \i /\sigma},t_3=\e^{2\pi\i\tau/\sigma}\big)$\;.};
  \path[commutative diagrams/.cd, |->, every label]
    (P0) edge node {} (P1)
    (P2) edge node {} (P0)
    (P1) edge node {} (P2);
\end{tikzpicture}\nn
\end{equation}
We can call this structure the $q$-Virasoro \textit{modular triple}, and it can be compactly encoded into the existence of three screening currents $\mb{S}_{12}(X)$, $\mb{S}_{23}(X)$, $\mb{S}_{31}(X)$ satisfying the relations (for all $i=1,2,3$)
\begin{align}
\left[\mb{T}_{m,i},\mb{S}_{12}(X)\right]=\left[\mb{T}_{m,i},\mb{S}_{23}(X)\right]=\left[\mb{T}_{m,i},\mb{S}_{31}(X)\right]=\textrm{total $X$-difference}\;,\nn
\end{align}
eventually leading to the existence of three commuting sets of Ward Identities that correlators of the modular triple must satisfy. It is clear that any two pairs among the above deformation parameters are related by the same $S\in \mathrm{SL}(2,\mathbb{Z})$ element as before, up to $q\leftrightarrow t^{-1}$ exchanges. Therefore, one can extract three different modular doubles out of the triple by simply neglecting a ``chiral" sector, in a cyclic order. Vice versa, the triple can be constructed by gluing three doubles upon identification of common ``chiral" sectors. This reasoning also reveals the need, origin and uniqueness of the $\mathrm{SL}(3,\mathbb{Z})$ structure: the $q\leftrightarrow t^{-1}$ exchange symmetry of the $q$-Virasoro algebra combines with the $\mathrm{SL}(2,\mathbb{Z})$ symmetry of the double and enhances to the $\mathrm{SL}(3,\mathbb{Z})$ symmetry in the triple, establishing full democracy among the $(\tau,\sigma)$ parameters which otherwise would be on a different footing. Hopefully, our algebraic construction can serve as a basis for a better understanding of the  triple factorization structure of the $\mathbb{S}^5$ partition function of supersymmetric Yang-Mills theories derived from localization, as well as for developing further the proposal of \cite{Lockhart:2012vp} (see also \cite{Hatsuda:2013oxa}) for the non-perturbative completion of refined topological strings \cite{Aganagic:2003db,Iqbal:2007ii,Awata:2008ed}. 

The rest of this paper is organized as follows. In section \ref{section:double}, we review the basics of the $q$-Virasoro algebra to set the notations, as well as the modular double construction. In section \ref{section:triple}, we fuse three modular doubles together and introduce the modular triple. In section \ref{section:formal-boson}, we give a 2d CFT-like construction of the modular triple, conjecturing for the first time a Lagrangian formulation for a $q$-Virasoro system, resembling a non-local version of ordinary Liouville theory. In section \ref{section:discussions}, we further our discussion  and outline possible applications and directions for future research. We supplement the paper with two appendices, where we collect the definition of special functions used in the main body and some technical computation.

\section{$q$-Virasoro and its modular double}\label{section:double}

\subsection{$q$-Virasoro algebra and notations}

We start by reviewing basics of the $q$-Virasoro algebra and its free boson realization through the Heisenberg algebra \cite{Shiraishi:1995rp,Awata:1995zk,frenkel:1997qW}. We will also introduce some notations for later convenience.

The \textit{$q$-Virasoro algebra} is the associative algebra generated by $\{\mb{T}_m, m \in \mathbb{Z}\}$\footnote{We will use straight letters in boldface, like $\mb{T}, \mb{a}, \mb{P, Q}$ and so, to denote operators.} with parameters $q, t \in \mathbb{C}$, $ p\equiv qt^{-1}$, satisfying the commutation relations 
\begin{align}
  \sum_{k \geq 0} f_k(\mb{T}_{m-k} \mb{T}_{n+k} - \mb{T}_{n-k} \mb{T}_{m+k}) = - \frac{(1-q)(1-t^{-1})}{1-p}(p^m - p^{-m}) \delta_{m+n,0} \ .
\end{align}
The constants $f_k$ are encoded into the series expansion of the structure function $f(z) \equiv \exp[\sum_{m > 0} \frac{(1-q^m)(1-t^{-m})}{m(1+p^m) }z^m]$, namely $f(z)=\sum_{k = 0}^{+\infty} f_k z^k$. The above commutation relations can be reorganized as
\begin{align}
 & f\left(\frac{w}{z}\right)\mb{T}(z)\mb{T}(w) - f\left(\frac{z}{w}\right)\mb{T}(w)\mb{T}(z) = - \frac{(1-q)(1-t^{-1})}{1-p}\left( \delta \left(p \frac{w}{z}\right) -  \delta \left(p^{-1} \frac{z}{w}\right)\right)\ ,
\end{align}
where we defined the $q$-Virasoro current $\mb{T}(z) \equiv \sum_{m \in \mathbb{Z}} \mb{T}_m z^{-m}$ and the multiplicative $\delta$ function $\delta(z)\equiv \sum_{m \in \mathbb{Z}}z^m$. We notice that the $q$-Virasoro algebra is invariant under $q \leftrightarrow t^{-1}$, which is also known as quantum $q$-geometric Langlands equivalence \cite{Kimura:2015rgi,Aganagic:2017smx}. Under this swapping $p$ is invariant, namely $p \leftrightarrow p$.

The $q$-Virasoro algebra admits a representation in terms of the Heisenberg algebra, where the latter is generated by oscillators $\{\mb{a}_m, m \in \mathbb{Z}_{\ne 0}\}$ and zero modes $\mb{P}, \mb{Q}$, with nontrivial commutation relations (here we follow the conventions of \cite{Awata:2010yy})
\begin{align}
 \big[\mb{a}_m, \mb{a}_n\big] &= -\frac{1}{m}(q^{m/2} - q^{-m/2})(t^{-m/2} - t^{m/2})C^{[m]}(p)\delta_{m+n, 0}\;, \qquad \big[\mb{P}, \mb{Q} \big] = 2\;,
\end{align}
where $C^{[m]}(p)=(p^{m/2} + p^{-m/2})$ is the deformed Cartan matrix of the $A_1$ algebra. The $q$-Virasoro current $\mb{T}(z)$ can be represented as
\begin{align}\label{stress-tensor-freeboson}
  \mb{T}(z) = \mb{Y}(p^{-1/2}z) + \mb{Y}(p^{1/2}z)^{-1}, \quad \mb{Y}(z) =  \exp \Bigg[\sum_{m \ne 0} \frac{\mb{a}_m  \ z^{-m}}{p^{m/2} + p^{- m/2}} \Bigg] \; q^{ \sqrt{\beta} \mb{P}/2} p^{ 1/2}\ ,
\end{align}
where  $\beta\in \mathbb{C}$ is such that $q^\beta = t$ and normal ordering of the modes is implicitly assumed. Note that the $q\leftrightarrow t^{-1}$ requires $\sqrt{\beta}\leftrightarrow -1/\sqrt{\beta}$ under this swapping.

A central object in $q$-Virasoro theory is played by the \textit{screening current} $\mb{S}_+(x)$. 
If we introduce the $\xi$-shift operator $\widehat{T}_\xi$ acting in the multiplicative notation on $x$ as
\begin{align}
\widehat{T}_\xi f(x)&= f(\xi x)\;,
\end{align}
the defining property of the screening current is  
\begin{align}\label{TS}
  \big[ \mb{T}_m, \mb{S}_+(x) \big] = \frac{\widehat T_q-1}{x}\left(\cdots\right) \qquad \Rightarrow \qquad \Big[ \mb{T}_m, \oint \d x \ \mb{S}_+(x) \Big] = 0 \;,
\end{align}
for a suitable choice of integration contour and where the dots represent some (uninteresting) operator. The contour integral of $\mb{S}_+(x)$ is called the \textit{screening charge}, and the $q$-Virasoro algebra can be indeed defined as the commutant of the screening charge in the Heisenberg algebra. Concretely, the screening current can be  realized by the following operator 
\begin{align}\label{standard-screening-current}
  \mb{S}_+(x) \equiv \exp\Bigg[ - \sum_{m \ne 0} \frac{\mb{a}_m \ x^{-m}}{ q^{m/2} - q^{- m/2}} +\sqrt{\beta}\mb{Q}\Bigg]\;\frac{\Theta(x \ q^{-\sqrt{\beta}\mb{P}}; q)}{\Theta(x ; q)\Theta(q^{-\sqrt{\beta}\mb{P}}; q)}\; .
\end{align}
We stress that the dependence on the momentum operator $\mb{P}$ is different from the usual definition in existing literature (which is simply $x^{\sqrt{\beta}\mb{P}}$),\footnote{Our redefinition is possible because $[\mb{T}(z),\mb{P}]=0$ in the free boson realization. Let $\mb{\Omega}(x)$ be a vertex operator containing only $\mb{P}$ such that $\mb{\Omega}(q x) = \mb{\Omega}(x)$. Then the replacement $\mb{S}(x)\to \mb{\Omega}(x)\mb{S}(x)$ does not spoil the property (\ref{TS}).} and this change will be useful later. A crucial observation is that the screening current above is not the only operator that commutes with $\mb{T}_m$. In fact, one can define a similar screening current $\mb{S_-}(x)$ and charge by using the $q \leftrightarrow t^{-1}$ symmetry, namely
\begin{align}
 \mb{S}_-(x) \equiv \exp\Bigg[ - \sum_{m \ne 0} \frac{\mb{a}_m \ x^{-m}}{t^{-m/2} - t^{m/2}} - \frac{\mb{Q}}{\sqrt{\beta}} \Bigg]  \;  \frac{\Theta(x \ t^{ - \sqrt{\beta}^{-1}\mb{P}}; t^{-1})}{\Theta(x ; t^{-1})\Theta(t^{-\sqrt{\beta}^{-1}\mb{P}}; t^{-1})}  \;  ,
\end{align}
with the property that
\begin{align}\label{TS}
  \big[ \mb{T}_m, \mb{S}_-(x) \big] = \frac{\widehat T_{t^{-1}}-1}{x}\left(\cdots\right) \qquad \Rightarrow \qquad \Big[ \mb{T}_m, \oint \d x \ \mb{S}_-(x) \Big] = 0 \;.
\end{align}

\subsection{$q$-Virasoro modular double}

We recall that in standard 2d free boson theory, or more general physical 2d CFTs, the degrees of freedom factorize into holomorphic and anti-holomorphic sectors, each having an independent copy of Virasoro symmetry algebra generated by $\{\mb{L}_m\;, m\in\mathbb{Z}\}$ and $\{\overline{\mb{L}}_m\;, m\in\mathbb{Z}\}$ respectively. However, the two chiral sectors do not decouple completely as they share the same zero modes. Concretely, the real free boson $\phi(z, \bar z)$ has on-shell mode expansion
\begin{align}\label{phizz}
  \phi(z, \bar z) &= \sum_{m \ne 0} a_m z^{-m} + \sum_{m \ne 0} \bar a_m \bar z^{-m} - P \ln z\bar z - Q~,
  \end{align}
  while for most of the purposes one can work with either one of the two chiral fields 
  \begin{align}\label{phiz}
  \phi(z)& \equiv \sum_{m \ne 0} a_m z^{-m} - P \ln z - Q \;,\nonumber\\
  \bar\phi(\bar z)& \equiv \sum_{m \ne 0} \bar a_m \bar z^{-m} - P \ln \bar z - Q \;,
\end{align}
and restore the physical dependence on $z$ and $\bar z$ successively. Upon quantization, $a_m$, $\bar a_m$ and $P, Q$ are promoted to operators. Exponential operators of the fundamental field $\phi(z, \bar z)$ are primary operators, and amongst them there are two with conformal dimension $(1,1)$
\begin{align}\label{VirS}
\mb{s}_\pm(z, \bar z) &\equiv \e^{\mp\sqrt{\beta^{\pm 1}}\phi(z, \bar z)}\;.
\end{align}
These operators satisfy the crucial relations
\begin{align}
  \big[\mb{L}_m, \mb{s}_\pm(z, \bar z)\big] = \partial_z \left(\cdots\right), \qquad \big[\overline{\mb{L}}_m, \mb{s}_\pm(z, \bar z)\big] = \partial_{\bar z}\left(\cdots\right)\;, \label{S-property}
\end{align}
and hence their holomorphic and anti-holomorphic components 
\begin{align}\label{VirSS}
\mb{s}_\pm(z) &\equiv \e^{\mp\sqrt{\beta^{\pm 1}}\phi(z)}\;,\qquad \overline{\mb{s}}_\pm(\bar z) \equiv \e^{\mp\sqrt{\beta^{\pm 1}}\bar\phi(\bar z)}\;
\end{align}
can be used to define the screening currents and charges of each copy of the Virasoro algebra. It is worth noting that the exchange symmetry $\sqrt{\beta}\leftrightarrow -1/\sqrt{\beta}$ is the analogous the of the $q \leftrightarrow t^{-1}$ symmetry of the $q$-Virasoro algebra, while the existence of two (holomorphic and anti-holomorphic) chiral sectors should be accounted by multiple $q$-Virasoro sectors. How this can consistently be done in a non-trivial way was explored in \cite{Nedelin:2016gwu}, and here we recall the basic ideas before generalizing that construction. 

Since the $q$-Virasoro algebra is as a deformation of the ordinary Virasoro algebra,  it is natural to seek the $q$-deformed counterpart of (\ref{S-property}), given two commuting $q$-Virasoro algebras. In doing so, we can consider (loosely speaking) deformations of the form
\begin{align}
  z \to x^{(\I)} = \e^{2\pi \i \frac{X}{\omega_1}}\;, \qquad \bar z \to x^{(\II)} = \e^{2\pi \i \frac{X}{\omega_2}}\;, \qquad \mb{L}_m \to \mb{T}^{(\I)}_{m}\;, \qquad \overline{\mb{L}}_m \to \mb{T}^{(\II)}_{m}\;,
\end{align}
where $\omega_{i=1,2}$ are two (possibly complex) generic\footnote{If not stated otherwise, throughout this paper we will assume the equivariant parameters to be generic, namely for any $i\neq j$, $\omega_i/\omega_j$ will be an irrational number which can be given an imaginary part if needed.} \textit{equivariant parameters} and $X$ a (possibly complex) coordinate. We stress that the ``chiral" variables $x^{(\I,\II)}$ are neither independent or complex conjugate to each other. Rather, they introduce two different periodicities in the $X$-direction. 
More precisely, let us consider two $q$-Virasoro algebras labeled by $\textrm{A} = \I, \II$ and with deformation parameters 
\begin{align}\label{parameter-pair}
q^{(\I)} &\equiv \e^{2\pi \i \frac{\omega_{12}}{\omega_1}}\ , \quad t^{(\I)} \equiv \e^{-2 \pi \i\frac{\omega_{3}}{\omega_1}}\ ,\qquad\qquad
q^{(\II)} \equiv \e^{2\pi \i \frac{\omega_{12}}{\omega_2}}\ , \quad t^{(\II)} \equiv \e^{-2 \pi \i\frac{\omega_{3}}{\omega_2}}\ ,
\end{align}
\begin{align}
\omega_{12} \equiv \omega_1 + \omega_2\;,\qquad \omega_3\equiv -\beta\omega_{12}\;,
\end{align}
which can be realized by two commuting Heisenberg algebras $\{\mb{a}^{(\textrm{A})}_{m},m\in\mathbb{Z}/\{0\}\}_{\textrm{A}=\I,\II}$ together with the common zero modes $\mb{P, Q}$. Notice that we can trade $\beta$ for the third equivariant parameter $\omega_3$, which is however on different footing w.r.t. $\omega_1$ and $\omega_2$ (for the moment). Then one can verify that the \textit{modular double screening current} $\mb{S}(X)$ given by\footnote{For later convenience, we have extracted a square root of $\e^{2\pi\i m}$, resulting in the $(-1)^m$ factor.} 
\begin{multline}\label{modular-double-screening-current}
  \mb{S}(X) \equiv \exp\Bigg[ - \sum_{m \ne 0} \frac{(-1)^m \ \mb{a}^{(\I)}_{m}\  \e^{-m\frac{2\pi\i}{\omega_1}X}}{\e^{\i\pi m \omega_{2}/\omega_1}-\e^{-\i\pi m\omega_{2}/\omega_1}} +\\
  - \sum_{m \ne 0} \frac{(-1)^m\ \mb{a}^{(\II)}_{m}\  \e^{-m\frac{2\pi\i}{\omega_2}X}}{\e^{\i\pi m \omega_{1}/\omega_2}-\e^{-\i\pi m\omega_{1}/\omega_2}}+\\
   +\sqrt{\beta}\mb{Q}+ \frac{2 \pi \i \omega_{12}\sqrt{\beta}\mb{P}}{\omega_1 \omega_2} X  \Bigg]\ ,
\end{multline}
satisfies the crucial relations
\begin{align}\label{TSmod}
  \big[ \mb{T}^{(\I,\II)}_{m}, \mb{S}(X) \big] &= \e^{-\frac{2\pi\i\omega_{12} X}{\omega_1\omega_2}}(T_{\lambda_{\I,\II}}-1)\left(\cdots\right)\;,\qquad \lambda_{\I,\II}=\omega_{2,1}\;,
  \end{align}
where we have defined the $\varepsilon$-shift operator $T_\varepsilon$ acting in the additive notation on $X$ as
\begin{align}
T_\varepsilon f(X)=f(X+\varepsilon)\;.
\end{align}
\textbf{Remark}. Notice that the screening current encodes two independent sets of oscillators, but the zero mode operators are shared, as in standard 2d CFT (\ref{phizz}). This is a more precise construction of the modular double compared with \cite{Nedelin:2016gwu}.

The oscillator part of the screening current is clearly factorized into two ``chiral" sectors, while it does not seem so for the zero mode part.\footnote{The naive splitting $\frac{\omega_{12}\sqrt{\beta}\mb{P}X}{\omega_1\omega_2}=\frac{\omega_{2}\sqrt{\beta}\mb{P}X}{\omega_1}+\frac{\omega_{1}\sqrt{\beta}\mb{P}X}{\omega_2}$ does not respect the $\omega_i$-periodicity in the $i$-th sector.} However, analogous to the holomorphic/anti-holomorphic splitting $\ln(z\bar z) = \ln z + \ln \bar z$, using (\ref{modTheta}) the momentum factor can also be factorized as 
\begin{align}
  \exp \Bigg[\frac{2\pi \i \omega_{12} \sqrt{\beta} \mb{P}}{\omega_1 \omega_2} X \Bigg] = \prod_{i=1,2}\frac{\Theta(\e^{\frac{2\pi\i}{\omega_i}X}\e^{-2\pi\i\frac{\omega_{12}}{\omega_i}\sqrt{\beta}\mb{P}};\e^{2\pi\i\frac{\omega_{12}}{\omega_i}})}{\Theta(\e^{\frac{2\pi\i}{\omega_i}X};\e^{2\pi\i\frac{\omega_{12}}{\omega_i}})\Theta(\e^{-2\pi\i\frac{\omega_{12}}{\omega_i}\sqrt{\beta}\mb{P}};\e^{2\pi\i\frac{\omega_{12}}{\omega_i}})}\;\e^{-\frac{\i \pi}{6}\big( \frac{\omega_{12}^2}{\omega_1 \omega_2} + 1 \big)} \;.
  \end{align}
Thanks to this factorization, one can extract the $\textrm{A}^{\textrm{th}}$ ``chiral" component $\mb{S}^{(\textrm{A})}(X)$ of the modular double screening current depending only on the ``chiral" variable $x^{(\textrm{A})}$
\begin{align}
\mb{S}^{(\textrm{A})}(X) &\equiv \exp\Bigg[ - \sum_{m \ne 0} \frac{\mb{a}^{(\textrm{A})}_{m} \ (x^{(\textrm{A})})^{ - m}}{(q^{(\textrm{A})})^{m/2} - (q^{(\textrm{A})})^{-m/2}}+\sqrt{\beta} \mb{Q}\Bigg]\frac{\Theta(x^{(\textrm{A})} (q^{(\textrm{A})})^{-\sqrt{\beta} \mb{P}};q^{(\textrm{A})})}{\Theta(x^{(\textrm{A})};q^{(\textrm{A})})\Theta((q^{(\textrm{A})})^{-\sqrt{\beta} \mb{P}};q^{(\textrm{A})})}\ ,
\end{align}
coinciding with (\ref{standard-screening-current}) upon the  identifications $q\sim q^{(\textrm{A})}$, $t\sim t^{(\textrm{A})}$, $\mb{a}_{m}\sim \mb{a}^{(\textrm{A})}_{m}$, $x \sim x^{(\textrm{A})}$ for fixed $\textrm{A}=\I,\II$, and to be compared with the holomorphic/anti-holomorphic components in (\ref{phiz}), (\ref{VirSS}).
  
The form of the dependence on $\mb{P}$ in (\ref{standard-screening-current}) is important for $\mb{S}_+(x)$ to descend from the screening current of the modular double as one of its ``chiral" components. The relation (\ref{TSmod}) immediately follows from this expression and (\ref{TS}). Remarkably, we can now consider not only the integrals of the ``chiral" components $\mb{S}^{(\textrm{A})}(X)$ defining the usual screening charges (as in the ordinary undeformed case), but also (and more interestingly) the integral of $\mb{S}(X)$ defining the screening charge of the modular double.

To summarize, two independent $q$-Virasoro algebras with deformation parameters coming in the specific form (\ref{parameter-pair}) related by $\textrm{SL}(2,\mathbb{Z})$ can be fused into what we called the  modular double, which can be denoted by $\mathcal{MD}(\omega_1,\omega_2;\beta)$. In the equivariant parametrization (\ref{parameter-pair}), the $\textrm{SL}(2,\mathbb{Z})$ symmetry simply amounts to the permutation symmetry $\omega_1\leftrightarrow\omega_2$. The whole structure can be encoded into the existence of a single screening current $\mb{S}(X)$ (\ref{modular-double-screening-current}) which simultaneously commutes (up to total differences) with two independent sets of $q$-Virasoro generators whose deformation parameters are related by $\textrm{SL}(2,\mathbb{Z})$.

\textbf{Remark}. The name modular double comes from the fact that the screening current $\mb{S}(X)$ shows manifest $\textrm{SL}(2,\mathbb{Z})$ symmetry. In fact, let $S \in \textrm{SL}(2,\mathbb{Z})$ act on certain  modular parameters $\tau,\sigma,\chi$ by 
\begin{equation}
\begin{tikzcd}[column sep=20pt]
(\tau,\sigma,\chi)\arrow[mapsto]{r}{S} & (-1/\tau,-\sigma/\tau,-\chi/\tau)
\end{tikzcd}\;.
\end{equation}
Then we can define $\tau \equiv \omega_{2}/\omega_1$, $\sigma\equiv -\omega_3/\omega_1$, $\chi\equiv X/\omega_1$ and $q^{(\I)} = \e^{2 \pi \i \tau}$, $t^{(\I)}=\e^{2\pi\i\sigma}$, $x^{(\I)} = \e^{2\pi \i \chi}$, and see that
\begin{equation}\label{taumod}
\begin{tikzcd}[column sep=20pt]
  \left(q^{(\I)} , t^{(\I)} ,x^{(\I)}  \right)\arrow[mapsto]{r}{S} &\left(q^{(\II)} , t^{(\II)},x^{(\II)} \right)
\end{tikzcd}\ .
\end{equation}
The parametrization in terms of the equivariant parameters $\omega_i$ is convenient because the $\textrm{SL}(2,\mathbb{Z})$ structure is simply reflected in the permutation symmetry $\omega_1\leftrightarrow\omega_2$.

Before ending this section, we observe that, among other things, the modular double has allowed us to consider two ``chiral" sectors resembling the holomorphic/anti-holomorphic structure of ordinary 2d CFTs. However, the analogous of the $\sqrt{\beta}\leftrightarrow-1/\sqrt{\beta}$ symmetry (\ref{VirS}) has now disappeared because the simple $q\leftrightarrow t^{-1}$ exchange of the ``chiral" theory is not allowed anymore. The reason is that in deriving the property (\ref{TSmod}), it is crucial to use the $\omega_{1,2}$-periodicity in the $\I,\II$ sectors, which is not respected by the equivariant parameter $\omega_3$ parametrizing $t^{(\I,\II)}$. It is therefore natural to ask whether it is possible to restore this symmetry by considering a third ``chiral" sector with $\omega_3$-periodicity  and such that there is complete democracy among the equivariant parameters $\omega_1$, $\omega_2$ and $\omega_3$. Remarkably, we give a positive answer to this question in the next section, the key being to combine the ``old" $q\leftrightarrow t^{-1}$ symmetry of each ``chiral" sector and the ``new" $\textrm{SL}(2,\mathbb{Z})$ symmetry of the modular double into an $\textrm{SL}(3,\mathbb{Z})$ action on $(\tau,\sigma)$.

\section{$q$-Virasoro modular triple}\label{section:triple}
In this section, we will generalize the above results and ``glue" multiple copies of modular doubles. Let us specify our conventions and introduce some useful notations. Since we will be dealing with multiple commuting copies of $q$-Virasoro algebras with different parameters, we will use lower case $i= 1, 2, 3$ to label them with natural cyclic identification $i + 3 \sim i$.\footnote{As we will see, three is the maximal number of copies in this paper, under a set of assumptions.} Generic equivariant parameters $\omega_{i=1, 2, 3}\in\mathbb{C}$ are frequently used to parametrize  the $q$-Virasoro parameters $q, t$ in different copies. We will find useful to also define the combinations 
\begin{align}
\omega_{i,i+1}\equiv\omega_i+\omega_{i+1}~,\qquad \omega\equiv\omega_1+\omega_2+\omega_3\;. 
\end{align}

\begin{figure}[t]
\centering
\includegraphics[width=0.8\textwidth]{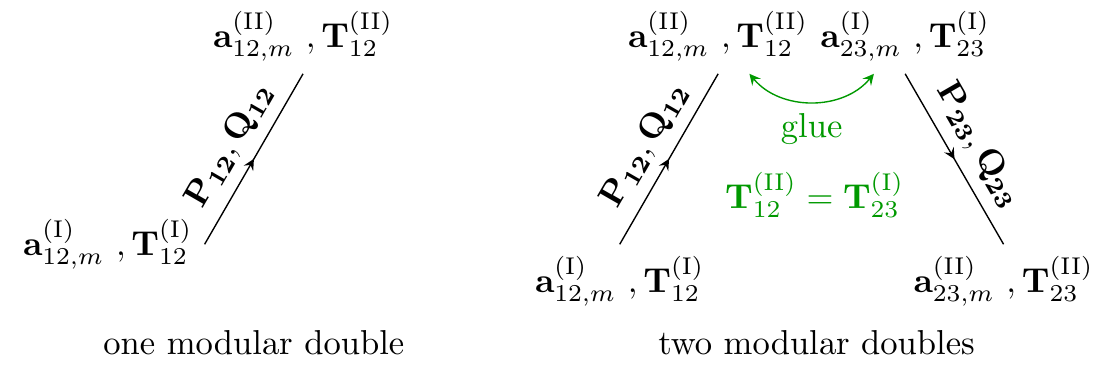}
\caption{A modular double is represented by an oriented edge, with the end-points representing the $\I,\II$ $q$-Virasoro sectors. Successive modular doubles are glued along a common vertex, where the generators are identified. The two screening currents associated to each vertex behave like $\mb{S}_\pm$ for the $q$-Virasoro algebra of that vertex.}\label{fig:gluing-modular-double}
\end{figure}

We start our discussion by ``non-trivially" gluing  two modular doubles and work out the consistency conditions among the $q,t$ parameters in different $q$-Virasoro algebras. The basic requirement is that all the $q$-Virasoro generators have to commute, up to total differences, with all the screening currents. As shown in Figure \ref{fig:gluing-modular-double}, we can conveniently represent a modular double by an oriented edge $(i,i+1)$ with the end-points $i$ and  $i+1$ representing the $\I$ and $\II$ $q$-Virasoro sectors respectively. In this notation, ``non-trivially" means that at the common vertex $i+1$, sector $\II$  of the $(i,i+1)$ double and sector $\I$  of the $(i+1,i+2)$ double are identified. At the level of the $q$-Virasoro generators meeting at that vertex, this amounts to impose
\begin{align}
\mb{T}^{(\II)}_{i,i+1}(z)=\mb{T}^{(\I)}_{i+1,i+2}(z) \ .
\end{align}

\begin{figure}[t]
  \centering
  \includegraphics[width=0.35\textwidth]{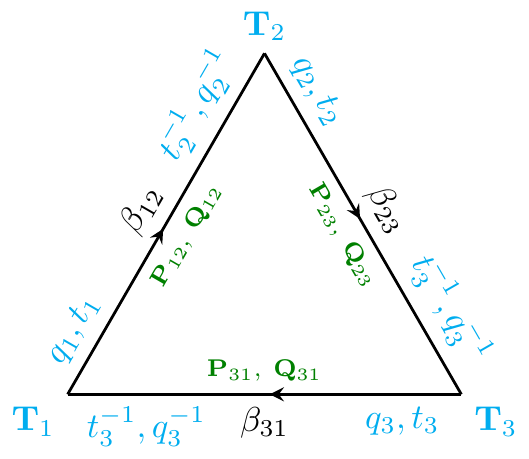}
  \caption{The result of gluing three modular doubles. For it to work, the parameters in the doubles must conspire in the form of (\ref{triple-parameters}). Each {\color{cyan}vertex} represents a $q$-Virasoro algebra, whose $q,t$ parameters can be chosen to be either of the adjacent blue ones; the two are related by $q \leftrightarrow t^{-1}$ leaving the algebra invariant. Each edge represents a modular double, which accepts those $q,t$-parameters along the edge, and a unique $\beta$ value. In the {\color{green!50!black}interior}, the zero modes in green are identified up to numerical constants.}
  \label{fig:modular-triple}
\end{figure}

For concreteness, let us start by considering two modular doubles $\mathcal{MD}(\omega_1,\omega_2;\beta_{12})$ and $\mathcal{MD}(\omega_2,\omega_3;\beta_{23})$ generated by the modes of the currents $\{\mb{T}^{(\textrm{A})}_{i,i+1}(z)\}^{\textrm{A}=\I,\II}_{i=1,2}$ respectively, with two special pairs of parameters
\begin{align}
  \Big(q^{(\I)}_{12}, t^{(\I)}_{12}; q^{(\II)}_{12}, t^{(\II)}_{12}\Big) &=  \Big(\e^ {2\pi \i \frac{\omega_{12}}{\omega_1}}, \e^ {2\pi \i \beta_{12}\frac{\omega_{12}}{\omega_1}}; \e^ {2\pi \i \frac{\omega_{12}}{\omega_2}}, \e^ {2\pi \i\beta_{12} \frac{\omega_{12}}{\omega_2}}\Big) \, \\
  \Big(q^{(\I)}_{23}, t^{(\I)}_{23}; q^{(\II)}_{23}, t^{(\II)}_{23} \Big) &=  \Big(\e^ {2\pi \i \frac{\omega_{23}}{\omega_2}}, \e^{2\pi \i \beta_{23}\frac{\omega_{23}}{\omega_2}}; \e^ {2\pi \i \frac{\omega_{23}}{\omega_3}}, \e^ {2\pi \i\beta_{23} \frac{\omega_{23}}{\omega_3}}\Big) \ .
\end{align}
Note that we have chosen the sectors labeled by $(\II)_{12}$ and $(\I)_{23}$, associated to vertex $2$, to have the same period $\omega_2$, which we are about to use for the gluing procedure. More concretely, we identify
\begin{align}\label{qtgluing}
\e^{2\pi\i\frac{\omega_{3}}{\omega_2}}=\e^{-2\pi\i\beta_1\frac{\omega_{12}}{\omega_2}} \ , \quad \e^{2\pi\i\beta_2\frac{\omega_{23}}{\omega_2}}= \e^{-2\pi\i\frac{\omega_{1}}{\omega_2}} \ ,
\end{align}
which, taking into account the $\omega_2$-periodicity, is also equivalent to
\begin{align}
q^{(\I)}_{23}=1/t^{(\II)}_{12} \ , \quad t^{(\I)}_{23} = 1/ q^{(\II)}_{12} \ , \quad p^{(\I)}_{23}=p^{(\II)}_{12}=\e^{2\pi\i\frac{\omega}{\omega_2}}\ ,
\end{align}
and
\begin{align}
\mb{a}^{(\I)}_{23,m} = \mb{a}^{(\II)}_{12,m}, \quad \omega_{12} \sqrt{\beta_1}\mb{P}_{12} = \omega_{23} \sqrt{\beta_2}\mb{P}_{23}\ , \quad \frac{\mb{Q}_{12}}{\omega_{12} \sqrt{\beta_{12}}} &= \frac{\mb{Q}_{23}}{\omega_{23} \sqrt{\beta_{23}}} \ ,
\end{align}
where the last identification ensures that $\big[\mb{P}_{12}, \mb{Q}_{12}\big] = \big[ \mb{P}_{23}, \mb{Q}_{23}\big] = 2$.
These conditions guarantee that $\mb{T}^{(\I)}_{23,m}= \mb{T}^{(\II)}_{12,m}$, which is made possible thanks to the $q \leftrightarrow t^{-1}$ invariance of the $q$-Virasoro algebra. See Figure \ref{fig:gluing-modular-double} for a visual illustration. Intuitively, from the viewpoint of $\mb{T}^{(\I)}_{23}$, $\mb{S}^{(\II)}_{12}$ behaves like $\mb{S}_-$ and thus it commutes with the former up to a total difference. More explicitly, we have the relations
\begin{align}
\big[ \mb{T}^{(\I)}_{12,m}, \mb{S}_{12}(X) \big] &= \e^{-\frac{2\pi\i\omega_{12}X}{\omega_1\omega_2}}(T_{\omega_2}-1)(\cdots)\;, & \big[ \mb{T}^{(\II)}_{12,m}, \mb{S}_{12}(X) \big] &=\e^{-\frac{2\pi\i\omega_{12}X}{\omega_1\omega_2}} (T_{\omega_1}-1)(\cdots)\ , \nn\\
\big[ \mb{T}^{(\I)}_{23,m}, \mb{S}_{23}(X) \big] &=\e^{-\frac{2\pi\i\omega_{23}X}{\omega_2\omega_3}} (T_{\omega_{3}}-1)(\cdots)\;,&\big[ \mb{T}^{(\II)}_{23,m}, \mb{S}_{23}(X) \big] &= \e^{-\frac{2\pi\i\omega_{23}X}{\omega_2\omega_3}}(T_{\omega_{2}}-1)(\cdots)\ , \nn\\   
\big[ \mb{T}^{(\II)}_{12,m}, \mb{S}_{23}(X) \big]  &= \e^{-\frac{2\pi\i\omega_{23}X}{\omega_2\omega_3}}(T_{\omega_3}-1)(\cdots)\;,& \big[ \mb{T}^{(\I)}_{23,m}, \mb{S}_{12}(X) \big] &=\e^{-\frac{2\pi\i\omega_{12}X}{\omega_1\omega_2}}(T_{\omega_1}-1)(\cdots)\ .
\end{align}
What remain to be checked are $\big[ \mb{T}^{(\I)}_{12,m}, \mb{S}_{23}(X) \big]$ and $\big[ \mb{T}^{(\II)}_{23,m}, \mb{S}_{12}(X) \big]$, which we postulate to vanish. The only non-trivial contributions to the commutators come from the zero modes
\begin{align}
\Big[\e^{\i\pi\frac{\omega_{12}\sqrt{\beta_{12}}\mb{P}_{12}}{\omega_1}},\e^{\sqrt{\beta_{23}}\mb{Q}_{23}}\Big]&=\e^{\sqrt{\beta_2}\mb{Q}_{23}} \e^{\i\pi\frac{\omega_{12}\sqrt{\beta_{12}}\mb{P}_{12}}{\omega_1}}\left(\e^{2\pi\i\beta_{23}\frac{\omega_{23}}{\omega_1}}-1\right)\;,\nn\\
\Big[\e^{\i\pi\frac{\omega_{23}\sqrt{\beta_{23}}\mb{P}_{23}}{\omega_3}},\e^{\sqrt{\beta_{12}}\mb{Q}_{12}}\Big]&=\e^{\sqrt{\beta_1}\mb{Q}_{12}} \e^{\i\pi\frac{\omega_{23}\sqrt{\beta_{23}}\mb{P}_{23}}{\omega_3}}\left(\e^{2\pi\i\beta_{12}\frac{\omega_{12}}{\omega_3}}-1\right)\;,
\end{align}
and their vanishing leads to the final constrains
\begin{align}
  \beta_1 = - \frac{\omega_3}{\omega_{12}} \ , \qquad \beta_2 =- \frac{\omega_1}{\omega_{23}}\ ,
\end{align}
consistent with the identifications anticipated in (\ref{qtgluing}). Now it is natural to glue more than two copies of modular doubles one after another, while requiring all generators to commute with all modular double screening currents up to total differences. This turns out to be quite restrictive, and \textit{at most three copies} of $q$-Virasoro algebras can be glued in the aforementioned way. Moreover, the third copy needs to be glued onto the first copy. More details about this rigidity can be found in appendix \ref{sec:gluing}. Pictorially, the three modular doubles labeled and represented by the edges $(i,i+1)=(12),(23),(31)$ have to form a closed triangle in which each vertex $i=1,2,3$ corresponds to one $q$-Virasoro algebra\footnote{We will take it to be the $\I$ copy in each double, so that we can neglect the index $\I,\II$.}, see Figure \ref{fig:modular-triple} for an illustration. More concretely, maximally three modular doubles  can be glued together
\begin{equation}
\begin{tikzpicture}[commutative diagrams/every diagram]
  \node (P0) at (0,-1) {$(12):  \mathcal{MD}(\omega_1,\omega_2; \beta_{12}\sim \omega_3)$\ ,};
  \node (P1) at (-4,1) {$(23):\mathcal{MD}(\omega_2,\omega_3; \beta_{23} \sim \omega_1)$} ;
  \node (P2) at (4,1) {$(31):\mathcal{MD}(\omega_3,\omega_1; \beta_{31} \sim \omega_2)$};
  \path[commutative diagrams/.cd, every arrow, every label]
    (P0) edge node {} (P1)
    (P2) edge node {} (P0)
    (P1) edge node {} (P2);
\end{tikzpicture}
\end{equation}
%\begin{equation}
%\begin{tikzpicture}[commutative diagrams/every diagram]
%  \node (P0) at (0,-1) {$(1):  \mathcal{MD}(q_1, t_1; t_2^{-1}, q_2^{-1}; \beta_1)$\;,};
%  \node (P1) at (-4,1) {$(2):\mathcal{MD}(q_2, t_2; t_3^{-1}, q_3^{-1}; \beta_2)$} ;
%  \node (P2) at (4,1) {$(3):\mathcal{MD}(q_3, t_3; t_1^{-1}, q_1^{-1}; \beta_3)$};
%  \path[commutative diagrams/.cd, every arrow, every label]
%    (P0) edge node {} (P1)
%    (P2) edge node {} (P0)
%    (P1) edge node {} (P2);
%\end{tikzpicture}
%\end{equation}
with the parameters of the $q$-Virasoro algebras at each vertex given in a fairly rigid form
\begin{align}
   q_i \equiv \e^{2\pi \i \frac{\omega_{i,i+1}}{\omega_i}}, \quad t_i \equiv \e^{ - 2\pi \i \frac{\omega_{i+2}}{\omega_i}} = q_i^{\beta_{i,i+1}}, \quad \beta_{i,i+1}\equiv - \frac{\omega_{i+2}}{ \omega_{i,i+1}} \label{triple-parameters}\;,
\end{align}
\begin{table}[t]
  \centering
  \begin{tabular}{c | c c c}
    Vertex $i$ & $1$ & $2$ & $3$\\
    \hline
    $q_i$ &  $\e^{2\pi \i \frac{\omega_1+\omega_2}{\omega_1} }$ & $\e^{2\pi \i \frac{\omega_2+\omega_3}{\omega_2} }$&$\e^{2\pi \i \frac{\omega_1+\omega_3}{\omega_3} }$  \\
    $t_i$  & $\e^{ - 2\pi \i \frac{\omega_3}{\omega_1} }$ & $\e^{- 2\pi \i \frac{\omega_1}{\omega_2} }$& $\e^{- 2\pi \i \frac{\omega_2}{\omega_3} }$\\
    $\beta_i$ & $- \frac{\omega_3}{\omega_1 + \omega_2}$ & $- \frac{\omega_1}{\omega_2 + \omega_3}$& $- \frac{\omega_2}{\omega_3 + \omega_1}$ 
  \end{tabular}
  \caption{Relations between $\omega_i$ and $(q_i, t_i)$ in various copies of $q$-Virasoro algebras.}\label{table:equivariant-params}
\end{table}as summarized also in Table \ref{table:equivariant-params}. The zero modes are identified up to numerical constants,
\begin{align}
  \omega_{i,i+1} \sqrt{\beta_{i,i+1}} \mb{P}_{i,i+1} &= \omega_{i+1,i+2}\sqrt{\beta_{i+1,i+2}} \mb{P}_{i+1,i+2}\;, & \frac{\mb{Q}_{i,i+1}}{\omega_{i,i+1}\sqrt{\beta_{i,i+1}}} &= \frac{\mb{Q}_{i+1,i+2}}{\omega_{i+1,i+2}\sqrt{\beta_{i+1,i+2}}} \;.
\end{align}
For this maximal case, we say that the three $q$-Virasoro algebras have fused into a \textit{modular triple}, and we can denote the three screening currents of the modular triple by $\mb{S}_{i,i+1}(X)$
\begin{multline}\label{Smodt}
\mb{S}_{i,i+1}(X)\equiv \exp\Bigg[ - \sum_{m \ne 0} \frac{(-1)^m \ \mb{a}^{(\I)}_{i,i+1,m} \ \e^{-m\frac{2\pi\i}{\omega_i}X}}{\e^{\i\pi m \omega_{i+1}/\omega_i} - \e^{-\i\pi m \omega_{i+1}/\omega_i}} +\\
- \sum_{m \ne 0}  \ \frac{(-1)^m \ \mb{a}^{(\II)}_{i,i+1,m} \ \e^{-m\frac{2\pi\i}{\omega_{i+1}}X}}{\e^{\i\pi m \omega_{i}/\omega_{i+i}} - \e^{-\i\pi m \omega_{i}/\omega_{i+1}}}  +\\
+\sqrt{\beta_{i,i+1}}\mb{Q}_{i,i+1} + \frac{2 \pi \i \omega_{i,i+1}\sqrt{\beta_{i,i+1}}\mb{P}_{i,i+1}}{\omega_i \omega_{i+1}} X \Bigg]\ .
\end{multline}

Let us summarize what we have proposed in this section. Given three commuting $q$-Vrasoro algebras with deformation parameters as in (\ref{triple-parameters}), one can construct three modular doubles by fusing any two of them, utilizing the $q \leftrightarrow t^{-1}$ invariance of each $q$-Virasoro algebra. Finally, the three modular doubles fuse in a circular fashion into one modular triple construction with properly chosen $\beta$'s and identifications of the zero modes. Also, it tuns out that no more than three copies can be glued in this way.
\vspace*{0.5em}

\textbf{Remark}. Each modular double $\mathcal{MD}(\omega_i,\omega_{i+1};\beta_i\sim \omega_{i+2})$ carries only an $\textrm{SL}(2,\mathbb{Z})$ structure realized in the $\omega_i\leftrightarrow \omega_{i+1}$ exchange symmetry, while the last equivariant parameter $\omega_{i+2}$ is on a different footing. The modular triple,  which we can denote by $\mathcal{MT}(\omega_1,\omega_2,\omega_3)$, establishes full democracy among all the equivariant parameters due to the manifest and complete permutation symmetry. In fact, each $\textrm{SL}(2,\mathbb{Z})$ structure is part of a more fundamental $\textrm{SL}(3,\mathbb{Z})$ structure. This is more manifest if we recall the parametrization (\ref{taumod}) for, say, the $i=1$ vertex of the modular triple
\begin{align}
q_1=\e^{2\pi\i\tau}\;,\quad t_1=\e^{2\pi\i\sigma}\;.
\end{align}
Then the other two vertices have deformation parameters $(q_2,t_2)$ and $(q_3,t_3)$ which are simply related to $(\tau,\sigma)$ by $\textrm{SL}(3,\mathbb{Z})$ transformations
\begin{align}
q_2&=\e^{-2\pi\i\sigma/\tau}\;,\quad t_2=\e^{-2\pi\i/\tau}\;,& q_3&=\e^{-2\pi\i/\sigma}\;,\quad t_3=\e^{2\pi\i\tau/\sigma}\;.
\end{align}
In this sense, the $q\leftrightarrow t^{-1}$ symmetry of each individual $q$-Virasoro algebra, together with the new $\textrm{SL}(2,\mathbb{Z})$ structure of the modular double, is enhanced to the $\textrm{SL}(3,\mathbb{Z})$ symmetry in the modular triple.

\section{Formal 2d CFT-like construction\label{section:formal-boson}}

In the previous section, we have shown that it is algebraically possible to fuse multiple $q$-Virasoro algebras when the deformation parameters are related by $\textrm{SL}(2,\mathbb{Z})$ or $\textrm{SL}(3,\mathbb{Z})$, and hence we have called the resulting constructions ($q$-Virasoro) modular double and triple respectively. These resemble ordinary structures of physical 2d CFTs, where the conformal algebra splits into holomorphic and anti-holomorphic parts giving rise to two (essentially independent) chiral sectors. It is therefore natural to ask whether our constructions admit a 2d CFT-like interpretation, possibly deriving some of the results from an action.

For any given $i \in \{1, 2, 3\}$ we will abbreviate the ordered sequence $(i, i+1, i+2)$ as $(i, j, k)$ to help shorten formulas. We also remind the cyclic identification $i+1\sim i$. Given three $q$-Virasoro algebras with parameters (\ref{triple-parameters}) in terms of the generic equivariant parameters $\omega_i\in\mathbb{C}$, we define the \textit{triple formal boson}
\begin{align}\label{triple-boson}
  \mb{\Phi}(X)& \equiv  \  \sum_{i=1}^3 \sum_{m \ne 0} \frac{\mb{h}_{i,m} \e^{\pi\i m\omega/\omega_i}\ \e^{-m\frac{2\pi\i}{\omega_i}X} }{{(1-\e^{2\pi\i m\omega_j/\omega_i})(1-\e^{2\pi\i m\omega_k/\omega_i})}} - \frac{\i\pi \mb{C}_2}{\omega_1\omega_2\omega_3} X^2 +\frac{  \mb{C}_1}{\omega_1\omega_2\omega_3} X+\mb{C}_0~,
  \end{align}
  with the operators $\mb{h}_{i,m}$, $\mb{C}_2$, $\mb{C}_1$, $\mb{C}_0$ satisfy the commutation relations (we display the non-trivial relations only)
  \begin{align}\label{hbos}
  [\mb{h}_{i,m},\mb{h}_{i,n}]&=-\frac{\e^{-\pi\i m\omega/\omega_i}}{m}(1-\e^{2\pi\i m\omega_j/\omega_i})(1-\e^{2\pi\i m\omega_k/\omega_i})\delta_{n+m,0}~,\qquad [\mb{C}_2,\mb{C}_1]=\hbar~,
\end{align}
where  $\hbar$ is a constant  (to be determined later). As we will see momentarily, this object is a useful and elegant device for compactly encoding the structure of the $q$-Virasoro modular double and triple. In order to show that, let us recall the definition of the shift operator $T_\varepsilon=e^{\varepsilon\partial_X}$ acting as
\begin{align}
T_\varepsilon f(X)&\equiv f(X+\varepsilon)~,
\end{align}
and the difference operator $d_{\varepsilon}$ defined as
\begin{align}
 d_{\varepsilon} f(X) &\equiv (T_{\varepsilon/2}-T_{-\varepsilon/2})f(X)=f(X + \varepsilon/2) - f(X - \varepsilon/2)~.
\end{align}
Then, let us define the bosons
\begin{multline} \label{double-boson} 
  \phi_{ij}(X)\equiv   d_{ \omega_{k}} \Phi(X)
  =\\
  - \sum_{m \ne 0}\Bigg[ \frac{(-1)^m\mb{h}_{i,m}\ \e^{-m\frac{2\pi\i}{\omega_i}X}}{\e^{\i\pi m\omega_j/\omega_i}-\e^{-\i\pi m\omega_j/\omega_i}}+\frac{(-1)^m\mb{h}_{j,m}\ \e^{-m\frac{2\pi\i}{\omega_j}X}}{\e^{\i\pi m\omega_i/\omega_j}-\e^{-\i\pi m\omega_i/\omega_j}}\Bigg]
-  \frac{2\pi\i \mb{C}_2}{\omega_i\omega_j}X+\frac{\mb{C}_1}{\omega_i\omega_j}~.
\end{multline}
The modes of these bosons resemble the ones appearing in the  screening currents of the modular triple (\ref{Smodt}), except that the we cannot readily identify the \textit{fundamental} oscillators $\mb{h}_{i,m},\mb{h}_{i+1,m}$ here with the \textit{root} oscillators $\mb{a}^{(\I)}_{i,i+1,m},\mb{a}^{(\II)}_{i,i+1,m}$ there, the commutation relations differing by the appearance of the deformed Cartan matrix $C^{[m]}(p)=\e^{\i\pi\omega m/\omega_i}+\e^{-\i\pi\omega m/\omega_i}$ of the $A_1$ algebra, which is non-trivial. The precise map is
\begin{align}\label{ahidentification}
\mb{a}^{(\II)}_{i-1,i,m}=\mb{a}^{(\I)}_{i,i+1,m}=\sqrt{-\i}(\e^{\pi\i m\omega /2\omega_i}+\i \e^{-\pi\i m\omega/2\omega_i})\mb{h}_{i,m}~,
\end{align}
which can be implemented by acting on $\phi_{ij}(X)$ with the bracket $(\omega/2,~)$ defined by 
\begin{align}
(\varepsilon, f(X))&\equiv \sqrt{-\i}(T_{\varepsilon/2}+\i T_{-\varepsilon/2})f(X)~.
\end{align}
Similarly, we can identify the zero modes $\mb{P}_{ij}$, $\mb{Q}_{ij}$ of the screening currents with $\mb{C}_2$, $\mb{C}_1$ as follows
\begin{align}
\sqrt{\beta_{ij}}\mb{P}_{ij}&=\frac{\sqrt{2}}{\omega_{ij}}\mb{C}_2~,\qquad \sqrt{\beta_{ij}}\mb{Q}_{ij}=-\frac{\sqrt{2}}{\omega_i\omega_j}(\mb{C}_1-\pi\omega\mb{C}_2/2)~,
\end{align}
implying also $\hbar=\omega_1\omega_2\omega_3$ so that $[\mb{P}_{ij},\mb{Q}_{ij}]=2$. Now, we can suggestively rewrite all the screening currents and charges of the modular triple in terms of the formal bosons $\Phi(X)$ or $\phi_{ij}(X)$ as
\begin{align}\label{screenings}
\mb{S}_{ij}(X)&= \e^{(\omega/2,\phi_{ij}(X))}\;,
\end{align}
resembling the usual 2d CFT definition (\ref{VirS}). We have focused on the construction of the screening currents because these are the most important operators for computational purposes. In appendix \ref{app:Tcurr}, we also discuss the construction of the $q$-Virasoro currents. 

Let us summarize what we have constructed so far: $i$) we have defined a formal boson $\Phi(X)$ (\ref{triple-boson}) which uniformly encodes three deformed ``chiral" bosons related by $\textrm{SL}(3,\mathbb{Z})$; $ii$) we have acted on $\Phi(X)$ with a simple difference operator $d_{\omega_k}$ to single out the bosons $\phi_{ij}(X)$ (\ref{double-boson}), each of which is closely related to a copy of the modular double; $iii$) we have used these fields to construct all the three screening currents of the  modular triple by introducing a sort of \textit{quantization} of the Killing form (\ref{screenings}). In view of the analogies with standard 2d CFT, in the rest of this section we will try to speculate on possible Lagrangian QFT constructions.

First of all, we would like to try to explain the origin of the three ``chiral" sectors encoded into $\Phi(X)$ and its mode expansion. We are thus led to postulate the equation of motion
\begin{align}\label{eom}
\Box_E\Phi(X)&=0~,\qquad \Box_E=\frac{1}{2\pi}\partial_X\prod_{i=1}^3 d_{\omega_i}~,
\end{align}
which can be derived from the (non-local) action 
\begin{align}\label{formalaction}
S_0[\Phi]&=\int \d X\; \partial_X d_{\omega_i} \Phi(X)d_{\omega_j} d_{\omega_k}\Phi(X)~,
\end{align}
where symmetrization in the indexes is left implicit. If we restrict to $X\in\mathbb{R}$ and $\omega_i\in\mathbb{R}$ such that the ratios $\omega_i/\omega_{i+1}$ are irrational $\forall i$ , solutions to (\ref{eom}) are of the form
\begin{align}
\Phi(X)&=\sum_{i=1,2,3}f_i(X)+p_3(X)~,
\end{align}
where $f_i(X)$ is a periodic function with period $\omega_i$ and $p_3(X)$ is a cubic polynomial. We can therefore match this on-shell field with the mode expansion of the formal boson in (\ref{triple-boson}), provided that we neglect the cubic coefficient of the zero mode polynomial 
\begin{align}
p_3(X)&=C_3 X^3- \frac{\i\pi C_2}{\omega_1\omega_2\omega_3} X^2 +\frac{ C_1}{\omega_1\omega_2\omega_3} X+C_0~.
\end{align} 
This can be justified, for example, by requiring an asymptotic behavior that forbids cubic terms. 

Next, we would like to quantize the theory and compute the 2-point function of the formal boson, using the ansatz (\ref{triple-boson}), (\ref{hbos}). As usual, let us postulate a vacuum state $| 0 \rangle$ annihilated by $\mb{h}_{i,m > 0}$, $\mb{C}_0$, and a dual vacuum state $\langle 0 |$  annihilated by $\mb{h}_{i, m < 0}$, with pairing $\langle 0| 0\rangle =1$. We can now easily compute the formal series 
\begin{align}
  \Delta_+(X) \equiv \langle 0 | \Phi(X) \Phi(0) | 0 \rangle &= - \ \sum_{i=1}^3 \sum_{m>0} \frac{\e^{\pi \i m \omega/\omega_i} \ \e^{-\frac{2\pi\i}{\omega_i}X}}{m (1 - \e^{2\pi \i m \omega_{j}/\omega_i}) (1 - \e^{2\pi \i m \omega_{k}/\omega_i})}~,\\
  \Delta_-(X) \equiv \langle 0 | \Phi(0) \Phi(X) | 0 \rangle &= - \ \sum_{i=1}^3 \sum_{m>0} \frac{\e^{\pi \i m \omega/\omega_i} \ \e^{\frac{2\pi\i}{\omega_i}X}}{m (1 - \e^{2\pi \i m \omega_{j}/\omega_i}) (1 - \e^{2\pi \i m \omega_{k}/\omega_i})}~.
\end{align}
In order to make more sense of these formal expressions, we have to consider an analytic continuation off the real line for $X$ and $\omega_i$. In this case we recognize that, when $\operatorname{Im}( \omega_i/\omega_{i+1}) \ne 0$ $\forall i$, $\Delta_\pm(X)$ are nothing but combinations of log of triple Sine functions and cubic Bernoulli polynomials (see \cite{Narukawa:2003} and appendix \ref{app:special})
\begin{align}
 \Delta_\pm (X) = & \ \log S_3(\omega/2\mp X|\vec\omega)\mp\frac{\i\pi}{6}B_{33}(\omega/2\mp X|\vec\omega)~,
\end{align} 
which are well-defined on the entire complex plane as long as $X$ does not hit any zero of the triple Sine function. We then define the propagator 
\begin{align}
G_E(X)&=\langle 0 |\textrm{T} \Phi(X) \Phi(0) | 0 \rangle=\left\{\begin{array}{ll}\Delta_+(X)~&~\textrm{ if }\textrm{Re}(X)\geq 0\\
\Delta_-(X)~&~\textrm{ if }\textrm{Re}(X)<0
\end{array}\right.~,
\end{align}
where $\textrm{T}$ denotes $X$-ordering. For consistency of the proposed action at the quantum level and of our quantization procedure, we would expect this propagator to be a Green's function of the kinetic operator $\Box_E$ arising from the equation of motion. As shown in appendix \ref{Green}, $G(X)\equiv G_E(-\i X)$ is in fact a Green's function for the ``Wick" rotated theory, namely 
\begin{align}\label{boxwick}
\Box G(X)&=\delta(X)~,\quad \Box=-\frac{\i}{2\pi}\partial_X\prod_{i=1}^3 d_{-\i\omega_i}~.
\end{align}
We speculate that a proper quantization of the proposed action would lead to the commutation relations (\ref{hbos}), which have been so far axiomatically given. Also, notice that the non-trivial part of the Green's function is captured by the $\log$ of the triple Sine function, as the kinetic operator annihilates any cubic polynomial. Such ambiguity is probably related to different boundary conditions or regularization prescriptions. 

Finally, by analogy with ordinary Liouville theory, we are led to propose the interacting theory
\begin{align}\label{Sint}
S_{\vec \mu}[\Phi]&=S_0[\Phi]+\sum_{i=1,2,3}\mu_i\int \d X\; \e^{\frac{2\pi\i\omega_{i,i+1}X}{\omega_i\omega_{i+1}}}\;\e^{(\omega/2,d_{\omega_i}\Phi(X))}\;,
\end{align}
where $\vec{\mu}\equiv(\mu_1,\mu_2,\mu_3)$ are coupling constants. At the quantum level, the $q$-Virasoro modular triple symmetry of the interacting model may be argued by considering the formal perturbative expansion of its correlation functions
\begin{align}\label{Spert}
\langle~ \cdots ~\rangle_{\vec\mu} = \sum_{n_1,n_2,n_3\geq 0}\langle ~\cdots~\prod_{i=1,2,3}\frac{\mu_i^{n_i}}{n_i !}\left(\int\d X\; \e^{\frac{2\pi\i\omega_{i,i+1}X}{\omega_i\omega_{i+1}}}\;\e^{(\omega/2,d_{\omega_i}\Phi(X))}\right)^{n_i}~ \rangle_{0}\;,
\end{align}
where $\langle ~\cdots ~\rangle_{\vec\mu}$ denotes the quantum v.e.v. in the theory described by $S_{\vec\mu}[\Phi]$. At any finite order in the coupling constants, the above quantity involve insertions of the modular triple screening charges in the free theory, and hence by construction it should satisfy three commuting copies of $q$-Virasoro constraints with deformation parameters related by $\textrm{SL}(3,\mathbb{Z})$, in the spirit of conformal matrix model technology \cite{Morozov:1994hh}.

\section{Discussion}\label{section:discussions}

In this note, we have reviewed the $q$-Virasoro modular double originally introduced in \cite{Nedelin:2016gwu}, and showed how it is possible to consistently fuse three copies of it into what we have called the $q$-Virasoro modular triple. Essentially, the $q\leftrightarrow t^{-1}$ symmetry of an individual $q$-Virasoro algebra and the $\textrm{SL}(2,\mathbb{Z})$ symmetry of the modular double can combined into the $\textrm{SL}(3,\mathbb{Z})$ structure of the modular triple. Under certain natural assumptions, we have also shown that no more than three copies can be glued in this way. Finally, we have given a 2d CFT-like construction of the modular triple and proposed for the first time a (non-local) Lagrangian formulation of a $q$-Virasoro system.

There are many directions worth pursuing for future research. For instance, it is natural to generalize the modular triple  to quiver $\textrm{W}_{q,t}$ algebras of \cite{Kimura:2015rgi}, including the elliptic case \cite{Nieri2017,Iqbal:2015fvd,Kimura:2016dys,Lodin:2017lrc}. Also, although the formal 2d CFT-like construction of the modular triple is rather elegant, it is incomplete. A more satisfactory understanding of the proposed non-local action for the proposed $q$-Virasoro model would be highly desirable, including a proper quantization procedure. New methods similar to the Ostrogradsky formalism \cite{GrosseKnetter:1993td} for higher order Lagrangians may be needed.

On the gauge theory side, we observe that  the \textit{time-extended} Nekrasov partition function of pure 5d $\textrm{U}(N)$ supersymmetric Yang-Mills theory can be written in terms of an infinite product of $q$-Virasoro screening charges \cite{Kimura:2015rgi}, naturally leading to $q$-Virasoro Ward Identities because of the defining relation (\ref{TS}). Since the Coulomb branch expression of the (squashed) $\mathbb{S}^5$ partition function can be written by gluing three copies of the $\mathbb{R}^4_{q,t}\times\mathbb{S}^1$  partition function in an $\textrm{SL}(3,\mathbb{Z})$ symmetric fashion, it is natural to expect that the time-extended $\mathbb{S}^5$ partition function satisfies the Ward Identities of the modular triple. However, if we consider a suitable linear combination of products of an arbitrary number of screening charges of the modular triple, we can formally obtain another solution to the same equations. If one insists on their identification, we obtain two different expressions of the same quantity. This reasoning supports the results of \cite{Pan:2016fbl} for the Higgs branch expression of the $\mathbb{S}^5$ partition function. In fact, the vigilant readers may have noticed the striking similarity between Figure \ref{fig:modular-triple} and the toric diagram of $\mathbb{S}^5$, represented as the $\mathbb{T}^3$ fibration over a filled triangle. Indeed, we observe some deeper connections between the two. In \cite{Nedelin:2016gwu}, it was shown that one can construct $\mathbb{S}^3$ partition functions out of the $q$-Virasoro modular double. We immediately notice that an $\mathbb{S}^3$ is represented by an edge of the $\mathbb{S}^5$ toric diagram, just like a modular double is an edge in Figure \ref{fig:modular-triple}. This analogy goes beyond just one $\mathbb{S}^3$: one can consider up to three \textit{intersecting} $\mathbb{S}^3$'s inside  $\mathbb{S}^5$, corresponding to the three edges in the diagram. It is not difficult to show \cite{FYM} that gauge theory partition functions on the intersecting $\mathbb{S}^3\cup \mathbb{S}^3 \cup \mathbb{S}^3\subset \mathbb{S}^5$ are naturally generated by free boson correlators involving an arbitrary number of screening charges of the modular triple (see also \cite{Nieri:2013vba,Gomis:2016ljm} for a related discussion in the context of 4d and 5d AGT). Hence, besides the close connection with three dimensional intersecting gauge theories, we also expect the modular triple construction to describe gauge theories on $\mathbb{S}^5$, which is in fact part of the motivation behind the proposal (\ref{Sint}), (\ref{Spert}) and the whole paper. We leave this topic for future work. More generally, one would expect that there exist other ``modular multiple" constructions that should be associated to gauge theories on $\mathbb{S}^3/\mathbb{Z}_k$ and their intersecting unions in other Sasaki-Einstein manifolds \cite{Qiu:2013aga,Qiu:2014oqa,Schmude:2014lfa}. It would be interesting to study the modular properties \cite{Shabbir:2015oxa,Qiu:2015rwp,Benvenuti:2016dcs} from this perspective.

Finally, it is worth noting that our results may represent a basis for developing further non-perturbative refined topological strings and large $N$ open/closed duality in the spirit of \cite{Lockhart:2012vp}, as well as for considering Little String Theories \cite{Witten:1995zh,Seiberg:1997zk,Losev:1997hx} at the origin of the $q$-deformation \cite{Aganagic:2015cta} from a novel perspective.

\acknowledgments
We thank Guglielmo Lockhart, Jian Qiu, Shamil Shakirov and J\"org Teschner for valuable discussions. We also thank the Simons Center for Geometry and Physics (Stony Brook University) for hospitality during the Summer Workshop 2017, at which some of the research for this paper was performed. M.Z. would like to thank Chamb\'o the French bulldog for the inspiration. 
The research of the authors is supported in part by Vetenskapsr\r{a}det under grant \#2014-5517, by the STINT grant and by the grant ``Geometry and Physics" from the Knut and Alice Wallenberg foundation.

\appendix

\section{Special functions}\label{app:special}

In this appendix, we recall the definition of several special functions which we use in the main body. Below, $r$ is a positive integer, and $\vec\omega\equiv(\omega_1, \ldots, \omega_r)$ is a collection of non-zero complex parameters. We frequently take $r = 2$ or 3 for concreteness. We refer to \cite{Narukawa:2003} for further details. 

The \textbf{multiple Bernoulli polynomials} $B_{r n}(Z|\vec\omega)$ are defined by the generating function
\begin{align}
  \frac{t^r \e^{Xt}}{\prod_{i=1}^r \e^{\omega_i t} - 1} = \sum_{m \ge \mathbb{N}} B_{rn}(X|\vec\omega) \frac{t^n}{n!}\;.
\end{align}
In particular, we use $B_{22}(X|\vec\omega)$ and $B_{33}(X|\vec\omega)$ in this note, and they are given explicitly by
\begin{align}
  B_{22}(X|\vec\omega) &=  \ \frac{X^2}{\omega_1 \omega_2} - \frac{\omega_1 + \omega_2}{\omega_1 \omega_2} X + \frac{\omega_1^2 + \omega_2^2 + 3 \omega_1 \omega_2}{6\omega_1 \omega_2}\;,\\
  B_{33}(X|\vec\omega) &= \ \frac{X^3}{\omega_1 \omega_2 \omega_3} - \frac{3(\omega_1 + \omega_2 + \omega_3)}{2\omega_1 \omega_2 \omega_3} X^2  + \frac{\omega_1^2 + \omega_2^2 + \omega_3^2 + 3 \omega_1 \omega_2 + 3 \omega_2 \omega_3 + 3 \omega_3 \omega_1}{2\omega_1 \omega_2 \omega_3}X+\nn\\
  & \qquad - \frac{(\omega_1 + \omega_2 + \omega_3)( \omega_1 \omega_2 +  \omega_2 \omega_3 + \omega_3 \omega_1)}{4\omega_1 \omega_2 \omega_3} \; .
\end{align}

The \textbf{$q$-Pochhammer symbols} are defined as
\begin{equation}
  (x;{q_1},\dots ,{q_r})_\infty \equiv \prod\limits_{{n_1},\ldots ,{n_r} = 0}^\infty  {(1 - x q_1^{{n_1}}\ldots q_r^{{n_r}})} \qquad {\text{when all }}|{q_i}| < 1\;.
\end{equation}
Other regions in the $q$-planes are defined through relations
\begin{equation}
  (x;{q_1},\dots ,{q_r})_\infty = \frac{1}{{(q_i^{ - 1}x;{q_1},\ldots ,q_i^{ - 1},...,{q_r})}}\;.
\end{equation}
They have exponentiated series expansion
\begin{equation}
  (x;{q_1},\dots ,{q_r})_\infty = \exp \left[ { - \sum\limits_{n > 0} {\frac{1}{{n(1 - q_1^n)\cdots(1 - q_r^n)}}{x^n}} } \right]\;.
\end{equation}

The \textbf{multiple Sine functions} $S_r(X|\vec\omega) $ can be defined by the $\zeta$-regularized product
\begin{align}
  S_r(X|\vec\omega) \sim \prod_{m_1, \ldots, m_r \in \mathbb{N}} \Big(X + \sum_{i=1}^r m_i \omega_i \Big)^{(-1)^{r+1}}\Big( - X + \sum_{i=1}^r (m_i + 1) \omega_i \Big)\;.
\end{align}
$S_r(X|\vec\omega)$ is symmetric in all $\omega_i$, has the reflection property $S_r(X|\vec \omega) = S_r(\omega - X|\vec \omega)^{(-1)^{r+1}}$ for $\omega\equiv \omega_1 + \ldots + \omega_r$, and the shift property
\begin{align}
  S_r(X + \omega_i|\vec\omega) = \frac{S_r(X|\vec\omega)}{S_{r-1}(X|\widehat{\vec \omega})} \ , \quad \widehat{\vec\omega} \equiv (\omega_1, \omega_{i-1},   \omega_{i+1}, \ldots, \omega_r)\ .\label{shift-property}
\end{align}

The \textbf{triple Sine function } $S_3(X|\vec\omega)$ has a useful factorization property. When $\operatorname{Im}(\omega_i/\omega_j) \ne 0$ for all $i \ne j$, $S_3(X|\vec\omega)$ factorizes
\begin{align}
  S_3(X|\vec\omega) = \e^{- \frac{\i\pi }{6} B_{33}(X|\vec\omega)} \prod_{i=1,2,3}\big( \e^{\frac{2\pi\i}{\omega_i}X}; q_i, t_i^{-1} \big)_\infty\;,
\end{align}
where $q_i$ and $t_i$ are given in (\ref{triple-parameters}) or table \ref{table:equivariant-params} in terms of $\omega_i$.

When $\operatorname{Re}\omega_i > 0$ and $\operatorname{Re}\omega > \operatorname{Re}X > 0$, the $S_3(X|\vec\omega)$ admits the integral representations
\begin{align}\label{intrep}
  \log S_3(X|\vec\omega) &= \int_{\mathbb{R} \pm \i 0} \frac{\d t}{t} \frac{\e^{X t}}{\prod_{i=1}^3 (e^{\omega_i t} - 1)}\mp\frac{\i\pi }{6} B_{33}(X|\vec\omega) \;.
\end{align}

The \textbf{$\Theta$ function} is defined as $\Theta (x;q) \equiv (x;q)_\infty(q{x^{ - 1}};q)_\infty$. Its modular properties are closely related to the $B_{22}(X|\vec \omega)$ polynomial
\begin{align}\label{modTheta}
  &\Theta (\e^{\frac{2\pi\i}{\omega_1}X};\e^{2\pi \i\frac{\omega _2}{\omega _1}})\Theta (\e^{\frac{2\pi\i}{\omega_2}X};\e^{2\pi \i\frac{\omega _1}{\omega _2}}) = \exp \Big[ { - \i\pi {B_{22}}(X|\vec\omega )} \Big]\;.
\end{align}
As one can construct linear functions of $X$ using $B_{22}(X|\vec\omega)$, one can factorize $\e^{(\ldots)X}$ in terms of products of $\Theta$ functions. Also, $q$-constants can be defined by using (products of) $\Theta$ functions, for instance
\begin{equation}
  {c_\alpha }(x;q) \equiv \frac{{\Theta ({q^\alpha }x;q)}}{{\Theta (x;q)}}{x^\alpha }\quad  \Rightarrow \quad {c_\alpha }(q x;q) = c(x;q)\;.
\end{equation}

\section{Green's functions}\label{Green}

In this appendix, we show that the logarithm of the triple Sine function can be viewed as a Green's function of the operator (\ref{boxwick}). Let us consider $G(X) \equiv \ln S_3(\omega/2+\i X| \vec \omega)$. Applying repeatedly (\ref{shift-property}), we have for $0 < \epsilon \ll 1$\footnote{It is crucial to introduce the regularization by $\epsilon$, as $1/\sin (\pi \omega^{-1}_1 \i X)$ is singular at $X = 0$, and we need to extract the difference between two singular functions.}
\begin{align}\label{distr}
  -\i \partial_X d_{-\i\omega_1 -\i \epsilon}d_{-\i\omega_2}d_{-\i\omega_3} G(X) = \pi \omega _1^{ - 1}\left[ {\frac{{\cos \pi \omega _1^{ - 1}(\i X + \epsilon )}}{{\sin \pi \omega _1^{ - 1}(\i X + \epsilon )}} - \frac{{\cos \pi \omega _1^{ - 1}(\i X - \epsilon )}}{{\sin \pi \omega _1^{ - 1}(\i X - \epsilon )}}} \right] \ , \quad \forall X \in \mathbb{C} \ .
\end{align}
\begin{figure}
  \centering
  \includegraphics[width=0.3\textwidth]{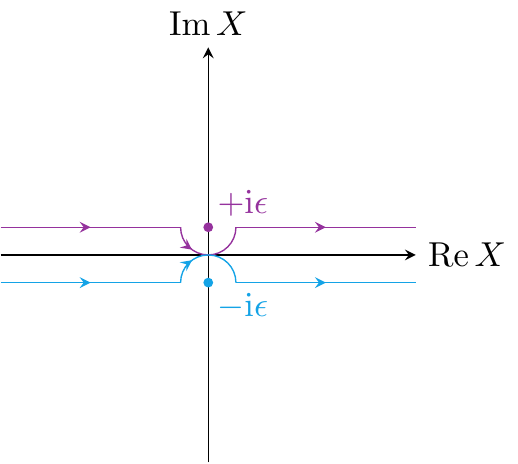}
  \caption{The integration contour when showing $G(X)$ is a Green's function of $\Box$.\label{fig:contour}}
\end{figure}Let us now specialize $X \in \mathbb{R}$, and observe that the two terms on the r.h.s. have a pole at $X = \pm \i \epsilon$  respectively. For any test function $f(X)$ that decays fast enough as $|X| \to +\infty$, we can compute its integral with the kernel (\ref{distr}) by residues, and in the $\epsilon\to 0^+$ limit we get
\begin{multline}
   \lim_{\epsilon \to 0^+}-\frac{\i}{2\pi}\int_\mathbb{R} \d X f(X) d_{-\i\omega_1 -\i \epsilon}d_{-\i\omega_2}d_{-\i\omega_3}\partial_XG(X)=\\
  =  \frac{\i\pi}{\omega_1} \lim_{\epsilon \to 0^+}
  \Bigg(
    \frac{1}{2}\mathop{\operatorname{Res}}\limits_{X \to + \i \epsilon }\frac{f(X)}{\sin \pi \omega_1^{-1}(\i X+\epsilon)} 
    - \bigg(-\frac{1}{2}\bigg)\mathop{\operatorname{Res}}\limits_{X \to - \i \epsilon }\frac{f(X)}{\sin \pi \omega_1^{-1}(\i X - \epsilon)} 
  \Bigg) 
  =   f(0)\;.
\end{multline}
Here, when evaluating the two integrals for $\pm \epsilon$, one can deform respectively the integration contours to $\mathbb{R} \pm \i \epsilon$ but going around the pole at $X = \pm \i \epsilon$ in a small half circle of radius $\epsilon$ from below/above, see Figure \ref{fig:contour}. In the second line, the principal values of the two integrals, \textit{i.e.} the integration in the regions $(- \infty \pm \i \epsilon, - \epsilon \pm \i \epsilon)$ and $(+ \epsilon \pm \i \epsilon, + \infty \pm \i \epsilon)$ cancel, while the two half-residues from the integration over the two half-circles add. Therefore, we conclude that for $X \in   \mathbb{R}$
\begin{align}
  \Box G(X) = \delta(X)\;.
\end{align}
In the above computation, we have just used the shift properties of the triple Sine function and a regularization prescription. Alternatively, we can take advantage of the integral representation (\ref{intrep}), and bringing the kinetic operator $\Box$ under the integral we immediately conclude that
\begin{align}
\Box G(X)=\frac{1}{2\pi}\int \d t\;\e^{\i X t}=\delta(X)\;.
\end{align}
Notice that, formally, the choice of different integration contours can lead to different Green's functions.

\section{Gluing multiple modular doubles}\label{sec:gluing}

We now show how to glue multiple modular doubles and derive the maximal number of modular doubles that can be glued. According to our recipe. Consider a collection of $r$ modular doubles labeled by $(i,i+1)$ with $i \in \mathcal{I}\equiv\{ 1, \ldots, r\}$, and Heisenberg algebras generated by operators $(\mb{a}^{(\I)}_{i,i+1,m},\mb{a}^{(\II)}_{i,i+1,m}, \mb{P}_{i,i+1}, \mb{Q}_{i,i+1})$ with equivariant parameters  
\begin{align}
  & (q_{i,i+1}^{(\I)}, t_{i,i+1}^{(\I)}) = (\exp \Big(2\pi \i \frac{\omega_{i,i+1}}{\omega_i}\Big), \exp \big(2\pi \i \frac{\beta_{i,i+1}\omega_{i,i+1}}{\omega_i}\big)) \ ,\nonumber \\
  & (q_{i,i+1}^{(\II)}, t_{i,i+1}^{(\II)}) = (\exp \Big(2\pi \i \frac{\omega_{i,i+1}}{\omega_{i+1}}\Big), \exp \big(2\pi \i \frac{\beta_{i,i+1}\omega_{i,i+1}}{\omega_{i+1}}\big)) \ .
\end{align}
Here $\omega_{i,i+1} \equiv \omega_i + \omega_{i+1}$. We now glue modular doubles consecutively. Consider the $(i,i+1)$ and $(i+1,i+2)$ modular doubles. We identify the generators $\mb{T}_{i,i+1,m}^{(\II)} = \mb{T}_{i+1,i+2,m}^{(\I)}$, which is translated to identification of the Heisenberg operators
\begin{align}
  & q_{i+1,i+2}^{(\I)} = 1/t_{i,i+1}^{(\II)}, \quad t_{i+1,i+2}^{(\I)} = 1/q_{i,i+1}^{(\II)} \ , \nonumber\\
  & \mb{a}_{i,i+1,m}^{(\II)} = \mb{a}_{i+1,i+2,m}^{(\I)} \ , \quad (q_{i,i+1}^{(\II)})^{\sqrt{\beta_{i,i+1}} \mb{P}_{i,i+1} /2} = (q_{i+1,i+2}^{(\I)})^{\sqrt{\beta_{i+1,i+2}} \mb{P}_{i+1,i+2} /2} \ .\label{constraints1}
\end{align}
By construction and the identification $\mb{T}_{i,i+1,m}^{(\II)} = \mb{T}_{i+1,i+2,m}^{(\II)}$ we achieve, for each $i = 1, \ldots, r - 1$, commutativity up to $\omega$-shifts, namely
\begin{align}
  \big[\mb{T}_{i,i+1,m}^{(\I)}, \mb{S}_{i,i+1}(X)\big] = & \ \e^{-\frac{2\pi\i\omega_{i,i+1}X}{\omega_i\omega_{i+1}}} (T_{\omega_{i+1}} - 1) (\ldots) \ , \quad \text{and same for }i \to i+1 \ ,\\
  \big[\mb{T}_{i,i+1,m}^{(\II)}, \mb{S}_{i,i+1}(X)\big] = & \ \e^{-\frac{2\pi\i\omega_{i,i+1}X}{\omega_i\omega_{i+1}}} (T_{\omega_{i}} - 1) (\ldots) \ , \quad \text{and same for }i \to i+1 \ ,\\
  \big[\mb{T}^{(\I)}_{i+1,i+2,m}, \mb{S}_{i,i+1}(X) \big] = & \ \e^{-\frac{2\pi\i\omega_{i,i+1}X}{\omega_i\omega_{i+1}}} (T_{\omega_{i}} - 1)(\ldots) \ , \nonumber\\
  \big[\mb{T}_{i,i+1,m}^{(\II)}, \mb{S}_{i+1,i+2} (X) \big] = &  \ \e^{-\frac{2\pi\i\omega_{i+1,i+2}X}{\omega_i\omega_{i+1}}} (T_{\omega_{i+1}} - 1)(\ldots) \ . \nonumber
\end{align}
This is of course not enough, and we impose full commutativity, \textit{i.e.}, $\forall i \in \mathcal{I}$
\begin{align}
  \big[\mb{T}_{j,j+1,m}^{(\I)}, \mb{S}_{i,i+1}(X)\big] = 0, \quad \forall j \in \mathcal{I}, \text{ and } j \ne i, i+1, \ \\
  \big[\mb{T}_{j,j+1,m}^{(\II)}, \mb{S}_{i,i+1}(X)\big] = 0, \quad \forall j \in \mathcal{I}, \text{ and } j \ne i, i-1 \ .
\end{align}
This leads to additional constrains on $\mb{P}_{i,i+1}$ and $\mb{Q}_{i,i+1}$,
\begin{align}
  \big[(q_{j,j+1}^{(\I)})^{ \pm \sqrt {\beta_{j,j+1}} \mb{P}_{j,j+1}/2},\e^{\sqrt {\beta_{i,i+1}} \mb{Q}_{i,i+1}} \big] = 0 , \quad \forall j \in \mathcal{I}, j \ne i, i+1, \nonumber\\
  \big[(q_{j,j+1}^{(\II)})^{ \pm \sqrt {\beta_{j,j+1}} \mb{P}_{j,j+1}/2},\e^{\sqrt {\beta_{i,i+1}} \mb{Q}_{i,i+1}} \big] = 0 , \quad \forall j \in \mathcal{I}, j \ne i, i-1 \ .  \label{constraints2}
\end{align}
\begin{figure}[t]
  \centering
  \includegraphics[width=0.5\textwidth]{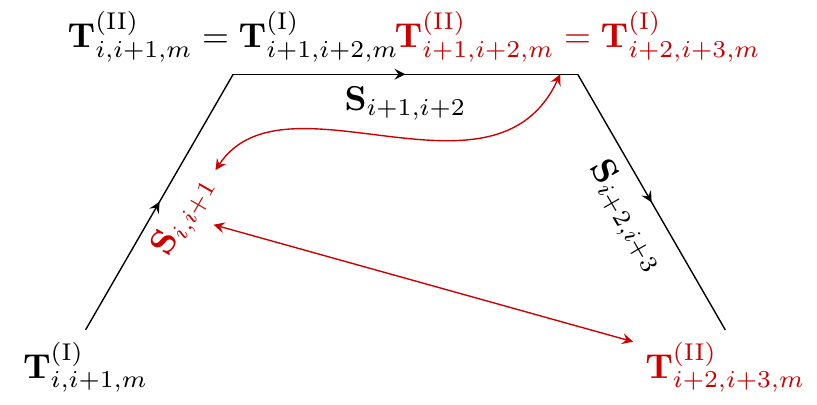}
  \caption{Gluing multiple modular double consecutively, where the red arrows indicate the origin of incompatible constraints.\label{fig:gluing-multiple-modular-double}}
\end{figure}Constraints (\ref{constraints1}) and (\ref{constraints2}) together lead to a set of equations involving undetermined integers $m^{(i)}, n^{(i)}$, $k^{(i)}$, $M^{(j,i)}$ and $N^{(j, i)}$
\begin{align}
  & \omega_{i,i+1} + m^{(i)}\omega_{i+1} = - \beta_{i+1,i+2} \omega_{i+1, i+2}\ , i = 1, \ldots, r-1  \ ,\\
  & \beta_{i,i+1}(\omega_{i,i+1}) + n^{(i)} \omega_{i+1} = - (\omega_{i+1, i+2}) \ , i = 1, \ldots, r-1 \ ,\\
  & \omega_{i, i+1} \frac{1}{2}\sqrt{\beta_{i,i+1}} \mb{P}_{i,i+1} + k^{(i)}\omega_{i+1} = \omega_{i+1, i+2} \frac{1}{2}\sqrt{\beta_{i+1,i+2}} \mb{P}_{i+1,i+2} \ , i = 1, \ldots, r-1 \ ,  \\
  & \frac{\omega_{j,j+1}}{\omega_j} \frac{1}{2} \sqrt{\beta_{j,j+1} \beta_{i,i+1}} \big[\mb{P}_{j,j+1}, \mb{Q}_{i,i+1}\big] = M^{(j,i)}, \quad \forall j = 1, \ldots, r, \text{ and } j \ne i, i+1 \ ,\\
  & \frac{\omega_{j,j+1}}{\omega_{j+1}} \frac{1}{2} \sqrt{\beta_{j,j+1} \beta_{i,i+1}} \big[\mb{P}_{j,j+1}, \mb{Q}_{i,i+1}\big] = N^{(j,i)}, \quad \forall j = 1, \ldots, r, \text{ and } j \ne i, i - 1 \ .
\end{align}

Let us analyze these constraints in detail. The equations in the third line relate the zero modes $\mb{P}_{i,i+1}$ in different modular double by nonzero proportionality, and therefore we conclude that $[\mb{P}_{i,i+1}, \mb{Q}_{j,j+1}] \ne 0$ for all $i, j$, simply because $[\mb{P}_{i,i+1}, \mb{Q}_{i,i+1}] = 2\neq 0$.

The equations in the last two lines come from (\ref{constraints2}) by applying the formula $\e^\mb{X}\e^\mb{Y} = \e^\mb{Y} \e^\mb{X} \e^{[\mb{X}, \mb{Y}]}$, valid when $[\mb{X}, \mb{Y}]$ commute with both $\mb{X}, \mb{Y}$, to $[\e^{\mb{X}}, \e^\mb{Y}] = \e^{\mb{Y}} \e^\mb{X}(\e^{[\mb{X}, \mb{Y}]} - 1)$. As we just saw, the integers $M^{(j, i)}, N^{(j, i)} \ne 0$. For any fixed $i$ and any fixed $j \ne i-1, i, i+1$, these two set of equations are incompatible for generic $\omega$'s, as they require
\begin{align}
  & \omega_{j,j+1} \frac{1}{2} \sqrt{\beta_{j,j+1} \beta_{i,i+1}} \big[\mb{P}_{j,j+1}, \mb{Q}_{i,i+1}\big] = M^{(j,i)} \omega_j = N^{(j,i)}\omega_{j+1} \ . 
\end{align}
In other words, if $r \ge 4$, one can always choose, for instance, $i=1$ and $j=3$, or, $i = 2$, and $j = 4$, and render the set of constraints unsolvable. See Figure \ref{fig:gluing-multiple-modular-double}. When $r = 3$, the constraints are solvable if and only if the three modular doubles are glued in a cyclic fashion, as in this special case, the constraints in the last two lines are absent.\footnote{In this cyclic case, we can formally set $r = 4$ and identify $i \sim i + 3$.} When $r = 2$, it is easy to find solutions as we presented in the main text. Finally, when $r = 3$ and the three modular doubles are glued cyclically, the constraints are solved by $m^{(i)} = n^{(i)} = -1$, $k^{(i)} = 0$, and
\begin{align}
  & \beta_{i,i+1} = - \frac{\omega_{i+2}}{ \omega_i + \omega_{i+1} } \ ,
  \end{align}
  \begin{align}
  & \omega_{i,i+1}\sqrt{\beta_{i,i+1}} \mb{P}_{i,i+1} = \omega_{i+1,i+2}\sqrt{\beta_{i+1,i+2}} \mb{P}_{i+1,i+2}, \quad  \frac{\mb{Q}_{i,i+1}}{\omega_{i,i+1}\sqrt{\beta_{i,i+1}}} = \frac{\mb{Q}_{i+1,i+2}}{\omega_{i+1,i+2}\sqrt{\beta_{i+1,i+2}}} \ .
\end{align}

\section{$q$-Virasoro currents}\label{app:Tcurr}
In this appendix, we show how it is possible to write the $q$-Virasoro current $\mb{T}^{(\I)}_{i,i+i}(x)$ in terms of the fundamental field $\Phi(X)$ introduced in section \ref{section:formal-boson}, where $x=\e^{2\pi\i X/\omega_i}$. Let us start by rewriting the operator (\ref{stress-tensor-freeboson}) as
\begin{align}
\mb{Y}^{(\I)}_{i,i+1}(x)&=\exp\Bigg[\frac{1}{2}\sum_\pm \sum_{m\neq 0} \frac{\mb{a}^{(\I)}_{i,i+1,m}\ \e^{-m\frac{2\pi\i}{\omega_i}(X+\omega/4)}}{\e^{\pi\i m\omega /2\omega_i}\pm \i \e^{-\pi\i m\omega /2\omega_i}}+\frac{\i\pi\omega_{i,i+1}\sqrt{\beta_{i,i+1}}\mb{P}_{i,i+1}}{\omega_i}+\frac{\i\pi\omega}{\omega_i}\Bigg] \ .
\end{align}
Now we introduce two fundamental fields $\Phi(X)$ and $\Phi(X)^\vee$ such that
\begin{align}
\mb{a}^{(\I)}_{i,i+1,m}=\sqrt{-\i}(\e^{\pi\i m\omega /2\omega_i}+\i \e^{-\pi\i m\omega/2\omega_i})\mb{h}_{i,m}=\sqrt{\i}(\e^{\pi\i m\omega /2\omega_i}-\i \e^{-\pi\i m\omega/2\omega_i})\mb{h}^\vee_{i,m}~.
\end{align}
We observe that when all the parameters are real, the ${}^\vee$ operation can be identified with Hermitian conjugation on $\mb{a}_{i,i+1,m}^{(\I)}$. This double possibility arises from the splitting $C^{[m]}(p)=\prod_\pm (p^{m/4}\pm \i p^{-m/4})$. This allows us to rewrite the above operator as
\begin{multline}
\mb{Y}^{(\I)}_{i,i+1}(x)=\exp\Bigg[\sum_{m\neq 0} \left(\frac{\sqrt{-\i}}{2} \mb{h}_{i,m}+\frac{\sqrt{\i}}{2}\mb{h}^{\vee}_{i,m} \right) \e^{-m\frac{2\pi\i}{\omega_i}(X+\omega/4)}+\\
+\frac{\i\pi\omega_{i,i+1}\sqrt{\beta_{i,i+1}}\mb{P}_{i,i+1}}{\omega_i}+\frac{\i\pi\omega}{\omega_i}\Bigg] \ ,
\end{multline}
or
\begin{align}
\mb{Y}^{(\I)}_{i,i+1}(x)=\exp\Bigg[\frac{\sqrt{-\i}}{2}T_{\omega_i/2}d_{\omega_j}d_{\omega_k}\Phi(X+\omega/4)+\frac{\sqrt{\i}}{2}T_{\omega_i/2}d_{\omega_j}d_{\omega_k}\Phi(X+\omega/4)^\vee+\frac{\i\pi\omega}{\omega_i}\Bigg] \ ,
\end{align}
where we also assumed $\mb{C}_2^\vee = -\mb{C}_2$.

\bibliographystyle{utphys}
%\bibliography{ref}

\providecommand{\href}[2]{#2}\begingroup\raggedright\endgroup

\end{document}